\documentclass[a4paper]{jpconf}

\usepackage[square,sort&compress,numbers]{natbib}
\usepackage{iopams}
\usepackage{amsmath}

\def\aap{Astron. Astrophys.}

\def\apj{Astrophys. J.}
\def\apjlett{Astrophys. J. Lett.}
\def\apjss{Astrophys. J. Suppl. Ser.}

\def\cjp{Can. J. Phys.}

\def\epjc{Eur. Phys. J. C}
\def\epjd{Eur. Phys. J. D}
\def\epl{Europhys. Lett.}

\def\ijmpa{Int. J. Mod. Phys. A}

\def\jcap{J. Cosmol. Astropart. Phys.}

\def\jpb{J. Phys. B}
\def\jpg{J. Phys. G}

\def\jms{J. Mol. Spectrosc.}

\def\mnras{Mon. Not. R. Astron. Soc.}
\def\nat{Nature}

\def\npb{Nucl. Phys. B}

\def\plb{Phys. Lett. B}

\def\pra{Phys. Rev. A}

\def\prc{Phys. Rev. C}
\def\prd{Phys. Rev. D}

\def\prl{Phys. Rev. Lett.}

\def\pscr{Phys. Scr.}

\def\rpp{Rep. Prog. Phys.}
\def\rmp{Rev. Mod. Phys.}
\def\sci{Science}

\newcommand{\E}[1]{\ensuremath{\times 10^{#1}}}
\newcommand{\yr}{\ensuremath{\textrm{yr}^{-1}}}

\begin{document}

\title{Astronomical and laboratory searches for space-time variation of fundamental constants}

\author{J. C. Berengut}
\address{School of Physics, University of New South Wales,
Sydney, NSW 2052, Australia}
\author{V. V. Flambaum}
\address{School of Physics, University of New South Wales,
Sydney, NSW 2052, Australia}
\address{European Centre for Theoretical Studies in Nuclear Physics and Related Areas (ECT*), Strada delle Tabarelle 286, I--38123, Villazzano (Trento), Italy}

\date{20 September 2010}

\begin{abstract}
We review recent activity searching for variations in the fundamental constants of nature in quasar absorption spectra and in the laboratory. While research in this direction has been ongoing for many decades, the topic has recently been stimulated by astronomical evidence for spatial variation in the fine-structure constant, $\alpha$. This result could be confirmed using different quasar data and atomic clock measurements, but there are significant challenges to obtain the required accuracy. We review existing measurements and discuss some of the most promising systems where any variations would be strongly enhanced.

\end{abstract}

\section{Introduction}
\label{sec:intro}

Recently, a highly significant ($\sim 4\sigma$) indication that there is a spatial gradient in values of the fine-structure constant, $\alpha$, was reported~\cite{webb10arxiv}. That is, in one direction on the sky $\alpha$ was larger in the past, while in the opposite direction it seems to have been smaller. This result has massive implications for the ``fine-tuning'' problem.  It is well known that the constants of nature are finely tuned to allow life to exist. For example, the production of carbon from helium in stars (the famous triple-$\alpha$ reaction) is sensitive to the position of a low-energy resonance, which in turn is very sensitive to variations in coupling constants. If the coupling constants were slightly different, there would be no resonance and hence no carbon-based life. There are many other examples. With the detection of spatial variation of coupling constants we begin to have a natural explanation of fine-tuning: we simply we appeared in the region of the Universe where values of fundamental constants are suitable for our existence.

We can only detect the variation of dimensionless fundamental constants. In this review we will discuss systems that are sensitive to variation of the fine-structure constant, $\alpha = e^2/\hbar c$, the proton $g$-factor, $g_p$, and the dimensionless mass ratios $\mu = m_e/m_p$ and $X_q=m_q/\Lambda_\textrm{QCD}$. Here $m_e$, $m_p$, and $m_q$ are the electron, proton, and light-current quark masses, respectively, and $\Lambda_\textrm{QCD}$ is the quantum chromodynamics scale, defined as the position of the Landau pole in the logarithm of the running strong coupling constant, $\alpha_s(r) \sim 1/\ln{(\Lambda_\textrm{QCD} r/\hbar c)}$. The proton mass $m_p$ is proportional to $\Lambda_\textrm{QCD}$ (if we neglect the few percent contribution of quark masses to the proton mass), therefore the relative variation of $\mu=m_e/m_p$ is approximately equal to the relative variation of $X_e=m_e/\Lambda_\textrm{QCD}$. In the Standard Model electron and quark masses are proportional to the vacuum expectation value of the Higgs field. In the same vein, $g_p$ is not a fundamental constant, but we can express its variation in terms of variation of light quark mass using the relationship $\delta g_p/g_p \approx -0.1\,\delta X_q/X_q$~\cite{flambaum04prd} (accurate calculations for nuclear $g$-factors can be found in \cite{flambaum06prc}).

The evidence for a cosmological spatial gradient in $\alpha$ motivates us to interpret other data in terms of spatial variation. However, many of the limits on variations are best interpreted as limits on fundamental constants other than $\alpha$, since the system being examined is more sensitive to these other constants. A minimal hypothesis is to expect all fundamental constants to vary along the same direction as the $\alpha$-gradient. There are some good theoretical justifications for this postulation. If the constants vary because they are coupled to a scalar field $\Phi$ that varies over space-time, for example the quintessence field $\Phi/c^2$ or a dimensionless dilaton field, then a fundamental constant, $X$, will vary according to
\begin{equation}
\label{eq:delphi}
\frac{\delta X}{X} = k_X \delta\Phi
\end{equation}
where $k_X$ is a dimensionless coupling coefficient. Then all constants will vary in the direction of the gradient of the scalar field $\Delta\Phi$, which in turn must be directed along the axis found by the $\alpha$-variation data in~\cite{webb10arxiv}.

Equation~\eref{eq:delphi} implies that the relative variation of different constants can be related to each other using equations like
\begin{equation}
\label{eq:k_mu_alpha}
k_\mu = R^\alpha_\mu\, k_\alpha
\end{equation}
where the constants $R^{X'}_X$ can be determined from observations and compared with theories of spatial variation. For example, applying Grand Unification of the interactions of the Standard Model to the relative variation of $\alpha$ and $\alpha_s$ (see, e.g.,~\cite{flambaum09ijmpa}) we find
\[
	\frac{\delta X_q}{X_q} \sim 35\, \frac{\delta\alpha}{\alpha}
\]
That is, the relative variation in $X_q$ may be much larger than the relative variation in $\alpha$. The coefficent here is model dependent, but large values are prevalent among models in which the variations come from high-energy scales. Therefore, variation of fundamental constants provides a unique probe of the predictions of Unification Theories (see, e.g.,~\cite{marciano84prl,langacker02plb,calmet02epjc,wetterich03jcap,dent03npb}).

In this review we discuss measurements of variation of fundamental constants in quasar absorption spectra and in atomic clocks. We interpret the astronomical measurements in the context of the observed spatial gradient in $\alpha$, relating other constants to this variation using expressions such as \eref{eq:k_mu_alpha}. Atomic clocks may prove to be the most reliable way to corroborate the detection of spatial $\alpha$-variation, and we discuss some of the proposed experiments where huge enhancements of the variation can be expected. There are many other systems which are not discussed here, including Big Bang Nucleosynthesis where hints of variation have been reported~\cite{dmitriev04prd,berengut10plb}. We refer the reader to recent reviews~\cite{flambaum09ijmpa,lea07rpp}.


\section{Quasar absorption spectra}
\label{sec:quasar}

We start our discussion with quasar absorption systems, since the remarkable recent detection of spatial variation in $\alpha$ from optical absorption spectra~\cite{webb10arxiv}, will provide us with a benchmark for all other studies.

\subsection{Optical atomic  spectra}
\label{sec:quasaralpha}

It is natural to analyse fine-structure intervals in the search for variation of $\alpha$. Measurements of $\alpha$-variation by comparison of cosmic and laboratory optical spectra were first performed by Savedoff as early as 1956~\cite{savedoff56nat}. There were numerous works successfully implementing this ``alkali-doublet'' method (see review~\cite{uzan03rmp}).

In 1999, a different approach was developed: the many-multiplet method~\cite{dzuba99prl,dzuba99pra}. It exploits the fact that relativistic corrections to atomic transition frequencies can exceed the fine-structure interval between the excited levels by an order of magnitude (for example, an $s$-wave electron does not have the spin-orbit splitting but it has the maximal relativistic correction to energy). We can express the $\alpha$-dependence of a transition frequency $\omega$ as
\begin{equation}
\label{eq:q}
\omega = \omega_0 + q x
\end{equation}
where
\begin{equation}
x = (\alpha/\alpha_0)^2 - 1 \approx 2\,\frac{\alpha - \alpha_0}{\alpha_0} ,
\end{equation}
$\alpha_0$ is the laboratory value of $\alpha$ and $\omega_0$ is the laboratory frequency of a particular transition. The coefficients $q$ can vary strongly from atom to atom and can have opposite signs in different transitions (for example, in $s$--$p$ versus $d$--$p$ transitions). Thus, any variation of $\alpha$ may be revealed by
comparing different transitions in different atoms in cosmic and laboratory
spectra. A statistical gain is also realised because many more spectral lines in different elements can be used. This method improves the sensitivity to any variation of $\alpha$ by more than an order of magnitude compared to the alkali-doublet method.

Relativistic many-body calculations are used to reveal the dependence of atomic frequencies on $\alpha^2$ (the $q$ coefficients). We have performed accurate many-body calculations of the $q$ coefficient for all transitions of astrophysical interest~\cite{dzuba99pra,dzuba99prl,dzuba03pra0,dzuba05pra,dzuba02praA,berengut04praB, berengut05pra,berengut06pra,dzuba08pra}. These are strong E1 transtions from the ground state in Mg\,I, Mg\,II, Ti\,II, Fe\,I, Fe\,II, Cr\,II, Ni\,II, Al\,II, Al\,III, Si\,II, Zn\,II, Mn\,II, as well as many other atoms and ions which are seen in quasar absorption spectra, but have not yet been used in the quasar measurements because of the absence of accurate UV transition laboratory wavelengths. For a ``shopping list'' of needed measurements, see Ref.~\cite{berengut09mmsait}.

From the very first analyses of quasar data that utilised the new method, hints of $\alpha$-variation were reported~\cite{webb99prl,webb01prl}. The largest analysis, with three independent samples of data containing 143 absorption systems spread over redshift range \mbox{$0.2 < z < 4.2$}, suggested that $\alpha$ was smaller in the past: \mbox{$\delta \alpha/\alpha = (-0.543 \pm 0.116) \times 10^{-5}$}~\cite{murphy03mnras}. However, these studies all used spectra taken at the Keck telescope. Similar studies from another group, using our method and calculations but a much smaller sample of data taken at the VLT, at first showed a stringent null constraints~(Srianand~\etal~\cite{srianand04prl}). More careful analysis of this sample suggested that the errors were underestimated by a large factor~\cite{murphy07prl,murphy08mnras,srianand07prl}.

The latest results, combining the Keck data and a new sample of 153 measurements from the VLT, shows a spatial variation in $\alpha$. This gradient, which we will refer to as the ``Australian dipole'', has a declination of around $-60^\circ$. This explains why the Keck data, restricted mainly to the northern sky since the telescope is in Hawaii at a latitude of 20$^\circ$~N, originally suggested a time-varying $\alpha$ that was smaller in the past. The VLT is in Chile, at latitude $25^\circ$~S, giving the new study much more complete sky coverage. The new results are entirely consistent with previous ones. The ``Australian dipole'' of $\alpha$-variation found by \cite{webb10arxiv} is
\begin{equation}
\label{eq:ausdipole}
	\frac{\delta\alpha}{\alpha_0} = (1.10\pm0.25)\E{-6}\, r \cos \psi
\end{equation}
where $\delta \alpha/\alpha_0 = (\alpha(\mathbf{r})-\alpha_0)/\alpha_0$ is the relative variation of $\alpha$ at a particular place $\mathbf{r}$ in the Universe (relative to Earth at $\mathbf{r}=0$). The function $r\cos\psi$ describes the geometry of the spatial variation: $\psi$ is the angle between the direction of the measurement and the axis of the Australian dipole, ($17.4\,(0.9)$~h, $-58\,(9)^\circ$) in equatorial coordinates. The distance function is the light-travel distance $r = ct$ measured in giga-lightyears. This is model dependent for large redshifts: the standard $\Lambda_\textrm{CDM}$ cosmology parametrized by WMAP5~\cite{hinshaw09apjss} is used to determine the light-travel time $t$. It is assumed here that $\delta\alpha/\alpha_0 = 0$ at zero redshift, which is supported by the data, however this assumption should be tested using the same absorption methods as are used at high redshift (e.g. by using absorbers within our own galaxy).

The new results of $\alpha$-variation are particularly striking in that if the VLT data and Keck data are considered separately, there is a high level of agreement between their dipole fits. In their paper~\cite{webb10arxiv}, the authors estimate the probability that the observed alignment between the independent Keck and VLT dipoles is due to chance at $\sim4\%$. Similarly, if one breaks the data into subsamples consisting of absorbers at $z<1.8$ and those at $z>1.8$ (approximately half the data in each subsample), then there is again agreement between the directions of the independent dipole axes.

In addition to the previously mentioned many-multiplet method results of Srianand~\etal from the small VLT sample~\cite{srianand04prl} (see also~\cite{chand04aap}), there are some high-accuracy single-absorption system $\alpha$-variation results using alkali-doublets (Si\,IV)~\cite{chand05aap} and single-ion (Fe\,II)~\cite{quast04aap,levshakov05aap,levshakov06aap,levshakov07aap} measurements. The alkali-doublet result suffers from the same problems as the small-sample many-multiplet result of~\cite{srianand04prl} (namely, sharp fluctuations in chi-squared vs. $\delta\alpha/\alpha$ graph which indicate failings in the chi-squared minimisation routine~\cite{murphy07prl}).

The Fe\,II results give stringent constraints in two absorbing systems: at $z=1.15$ towards HE0515--4414~\cite{levshakov06aap} and at $z=1.84$ towards Q1101--264~\cite{levshakov07aap}. It is interesting to compare the results of these analyses with the variation expected in these systems if the Australian dipole result is correct. Using \eref{eq:ausdipole}, we obtain an expected variation of $\delta\alpha/\alpha = (1.9\pm1.5)\E{-6}$ for the former, which was measured to be $(-0.07\pm0.84)\E{-6}$~\cite{levshakov06aap}. The absorber towards Q1101--264 has a larger expected variation of $\delta\alpha/\alpha = (3.8\pm2.0)\E{-6}$, and was measured to be $(5.4\pm2.5)\E{-6}$~\cite{levshakov07aap}. The single system results are therefore seen to be consistent with the Australian dipole.

\subsection{Molecular rotational quasar spectra}

Limits on variation of $\mu$ at high redshift can be obtained by comparison of different rotational-electronic excitations in molecular hydrogen. Taking advantage of newly available H$_2$ wavelengths, Ref.~\cite{reinhold06prl} reported a non-zero cosmological variation of $\mu$ in quasar spectra using molecular hydrogen transitions in the Ly-$\alpha$ forest. The authors obtained $\delta \mu/\mu = (2.4 \pm 0.6)\E{-5}$ at redshifts $z \approx 2.6 - 3.0$, i.e. a decrease of $\mu$ in the past 12 Gyr at the $3.5\sigma$ confidence level. 

More recent determinations, using the laboratory wavelengths and sensitivity coefficients presented in \cite{reinhold06prl,ubachs07jms,salumbides08prl} and carefully controlling systematics and accounting for known calibration errors in the VLT, obtain a more stringent constraint: $\delta\mu/\mu = (3.4\pm2.7)\E{-6}$~\cite{king08prl,malec10mnras} from four quasar absorption systems.

In the context of an observed dipole in $\alpha$-variation~\cite{webb10arxiv}, it makes sense to check whether the (limited) data supports interpretation as a spatial variation. In Ref.~\cite{berengut10arxiv1} the data was shown to support a statistically significant spatial variation of $\mu$ aligned with the Australian dipole, although it was noted that the paucity of data prevents any firm conclusion.

\subsection{Comparison of hydrogen hyperfine and UV transitions}

A comparison of the 21-cm hyperfine transition in atomic hydrogen with UV metal lines was performed for 9 quasar spectra with redshifts $0.23 \le z \le 2.35$~\cite{tzanavaris05prl,tzanavaris07mnras}. The ratio of the transition frequencies is proportional to the parameter $x=\alpha^2 \mu g_p$, and this was constrained to
\begin{equation}
\label{x_var}
\delta x/x = (6.3\pm 9.9)\times 10^{-6}\ .
\end{equation}
It was found that there was much more scatter in the data than would be expected from the statistical errors alone~\cite{tzanavaris07mnras}. In principle this could mean that the model is wrong, e.g. a spatial variation should be considered rather than a time variation. However, even the best fit dipole model (which does not align with the Australian dipole) cannot explain the scatter~\cite{berengut10arxiv1}. It is probable that velocity offsets between the hydrogen and metal absorbers cause dominating systematics. Two new measurements comparing H\,I 21-cm with UV lines in neutral carbon in single absorption systems have reported slightly stronger constraints in $x$~\cite{kanekar10apjlett}; these are shown in the first two lines of \tref{tab:singles}.

\subsection{Comparisons involving hyperfine and molecular rotational transitions}

Individual measurements of fundamental constant variation in quasar absorption spectra can be compared with the observed spatial $\alpha$-variation~\cite{berengut10arxiv1}. That is, we can regard the $\alpha$-variation, according to \eref{eq:ausdipole} with $r$ and $\psi$ calculated for the object, as the ``expected'' variation and compare other measurements with it (this is presented in \tref{tab:singles}).

\begin{table}[t]
\caption{\label{tab:singles} Comparison of expected variation, given by \eref{eq:ausdipole}, and measured variation of fundamental constants in different quasar absorption systems.
$R^\alpha_{g_p}$ and $R^\alpha_\mu$ are defined by equations like \eref{eq:k_mu_alpha}. As noted in Section~\ref{sec:intro}, $R^\alpha_{g_p}\approx -0.1\,R^\alpha_{q}$ and $R^\alpha_\mu \approx R^\alpha_q$ where $R^\alpha_q$ is the variation of light quark mass $X_q$ relative to $\alpha$-variation.
Errors in the expected $\alpha$-variation (the prefactors in the third column) are of the order $\sim1.5\E{-6}$.
}
\begin{tabular}{llccc}
\hline
System & Constant & Expected variation & Measurement & Ref. \\
       &          &$(\times 10^{-6})$ & $(\times 10^{-6})$ \\
\hline
\mbox{H\,I} + \mbox{C\,I} & $\alpha^2 \mu g_p$
    & $1.12\,(2 + R^\alpha_\mu + R^\alpha_{g_p})$
    & $6.64 \pm 0.84_\textrm{stat} \pm 6.7_\textrm{sys}$ & \cite{kanekar10apjlett} \\
 &
    & $-5.20\,(2 + R^\alpha_\mu + R^\alpha_{g_p})$
    & $7.0 \pm 1.8_\textrm{stat} \pm 6.7_\textrm{sys}$ & \cite{kanekar10apjlett} \\
\mbox{H\,I} + mol. rot. & $\alpha^2 g_p$
    & $0.50\,(2 + R^\alpha_{g_p})$
    & $-2.0 \pm 4.4$ & \cite{murphy01mnrasD} \\
 &
    & $-5.47\,(2 + R^\alpha_{g_p})$
    & $-1.6 \pm 5.4$ & \cite{murphy01mnrasD} \\    
\mbox{H\,I} + OH & $(\alpha^2/\mu)^{1.57} g_p$
    & $-1.04\,(3.14 - 1.57 R^\alpha_\mu + R^\alpha_{g_p})$
    & $4.4 \pm 3.6_\textrm{stat} \pm 10_\textrm{sys}$ & \cite{kanekar05prl} \\
OH & $(\alpha^2/\mu)^{1.85} g_p$
    & $0.50\,(3.70 - 1.85 R^\alpha_\mu + R^\alpha_{g_p})$
    & $-11.8 \pm 4.6$ & \cite{kanekar10apjlett0} \\
NH$_3$ & $\mu$
    & $-5.47\,R^\alpha_\mu$
    & $< 1.8\ (2\sigma)$ & \cite{murphy08sci} \\
 &
    & $1.34\,R^\alpha_\mu$
    & $< 1.4\ (3\sigma)$ & \cite{henkel09aap} \\
\hline
\end{tabular}
\end{table}

The frequency of the hydrogenic hyperfine line is proportional to $\alpha^2\mu g_p$; molecular rotational frequencies are proportional to $\mu$. Comparison of the two placed limits on variation of the parameter $\alpha^2 g_p$~\cite{drinkwater98mnras} in two quasar absorption spectra. A similar analysis was repeated using more accurate data for the same objects~\cite{murphy01mnrasD}, at $z=0.247$ and at $z=0.6847$, resulting in the limits shown in lines 3 and 4 of \tref{tab:singles}, respectively. The object at $z=0.6847$ is associated with the gravitational lens toward quasar B0218+357 and corresponds to lookback time $\sim 6.2$ Gyr.

Comparison of OH 18-cm with H$\,$I 21-cm, and different conjugate-satellite OH 18-cm lines, can be used to measure the combinations of fundamental constants $(\alpha^2/\mu)^{1.57}g_p$ and $(\alpha^2/\mu)^{1.85}g_p$, respectively. Resulting measurements are shown in lines 5 and 6 of \tref{tab:singles}.

\subsection{Enhancement of variation of $\mu$ in the inversion spectrum of ammonia}
\label{sec:NH3}

In 2004, van Veldhoven \etal\ suggested using a decelerated molecular beam of ND$_3$ to search for the variation of $\mu$ in laboratory experiments\cite{van-veldhoven04epjd}. The ammonia molecule has a pyramidal shape and the inversion frequency depends on the exponentially small tunneling of three hydrogen (or deuterium) nuclei through the potential barrier. Because of that, it is very sensitive to any changes of the parameters of the system, and particularly to the reduced mass for this vibrational mode.

High precision data on the redshifts of NH$_3$ inversion lines exist for the previously mentioned object B0218+357 at $z=0.6847$~\cite{henkel05aap}. Comparing them with the redshifts of rotational lines of CO, HCO$^+$, and HCN molecules from Ref.~\cite{combes97apj} allows strong limits on variation of $\mu$, and hence $X_e$, to be obtained~\cite{flambaum07prl}. The most accurate measurements, utilising new rotational spectra, give very stringent limits on $\mu$-variation in this quasar absorption system and the object PKS1830--211 at $z=0.8858$ (last two lines of \tref{tab:singles}).

\section{Clocks}
\label{sec:clocks}

\subsection{Optical atomic clocks}

Atomic clocks can be used to measure time-variation of fundamental constants in the Earth frame. Different optical atomic clocks utilize transitions that have positive, negative or small contributions of the relativistic corrections to frequencies, so comparison of these clocks can be used to measure $\alpha$-variation. The same methods of relativistic many-body calculations used in the quasar absorption studies can be used to calculate the dependence on $\alpha$ of different clocks~\cite{dzuba99pra,dzuba00pra, dzuba03pra0,angstmann04pra,angstmann04pra0,dzuba08pra0}. A summary of results is presented in Ref.~\cite{flambaum09cjp}. The relativistic effects are proportional
to $(Z\alpha)^2$, therefore the $q$ coefficients for optical clock transitions  may be substantially larger than in cosmic transitions since the clock transitions are often in heavy atoms (Hg\,II, Yb\,II, Yb\,III, etc.) while cosmic spectra contain mostly light atoms ($Z\lesssim 33$).

The temporal variation of $\alpha$ measured in the laboratory can be compared with the variation expected from the observed spatial gradient in $\alpha$~\cite{webb10arxiv} because the solar system moves along the axis of the dipole. The expected variation from the dipole was calculated to be~\cite{berengut10arxiv0}
\begin{equation}
\label{eq:clock_req}
\dot\alpha/\alpha = 1.35\E{-18} \cos\psi~\yr
\end{equation}
where $\psi$ is the angle between the motion of the Sun and the dipole. The best fit from~\cite{webb10arxiv} gives $\cos\psi \sim 0.07$ but this has an uncertainty of $\sim 0.15$. The $\alpha$-variation is modulated by the annual motion of the Earth around the Sun, $\delta\alpha/\alpha = 1.4\E{-20} \cos\omega t$~\cite{berengut10arxiv0} where $\omega$ refers to the angular frequency of the yearly orbit.

The current best constraint on time-variation of $\alpha$ was achieved by precisely measuring the frequency ratio of Hg\,II and Al\,II clocks several times over the course of a year~\cite{rosenband08sci}. Using our calculations, the rate of change of $\alpha$ is measured at \mbox{$\dot\alpha/\alpha = (-1.6\pm 2.3)\E{-17}~\yr$}. This limit will need to be improved by around two orders-of-magnitude in order to confirm or contradict the observed spatial gradient in $\alpha$. On the other hand, the quasar observations do not exclude time-variation of $\alpha$ below the rate of $\sim 10^{-16}~\yr$, and here the current laboratory limits are already competitive.

\subsection{Enhanced effect of $\alpha$-variation in Dy atom}

The sensitivity required by \eref{eq:clock_req} may be obtained by finding systems where $\alpha$-variation is strongly enhanced. Transitions between two almost degenerate levels in Dy atom can give a very high relative enhancement because these levels move in opposite directions if $\alpha$ varies~\cite{dzuba99prl,dzuba03pra0,dzuba08pra0}. The relative variation may be presented as $\delta \omega/\omega=K \delta \alpha /\alpha$ where the coefficient $K$ exceeds $10^8$ ($q=30\,000$ cm$^{-1}$, $\omega \sim 10^{-4}$ cm$^{-1}$). The values of $K=2 q/\omega$ are different for different hyperfine components and isotopes because $\omega$ changes. An experiment is currently underway to place limits on $\alpha$ variation using this transition~\cite{nguyen04pra,cingoz07prl}, however one of the levels has  quite a large linewidth and this limits the accuracy. The current limit is $\dot{\alpha}/\alpha=(-2.7 \pm 2.6) \times 10^{-15}$~yr$^{-1}$. Several other enhanced effects of $\alpha$ variation in atoms have been calculated~\cite{dzuba05pra0,angstmann06jpb}.

\subsection{Enhanced effects of $\alpha$-variation in highly charged ions}

Sensitivity to $\alpha$-variation increases with ion charge as $(Z_i+1)^2$~\cite{berengut10prl}. The most sensitive atomic systems will maximize the contributions from three factors: high nuclear charge $Z$, high ionization degree, and significant differences in the configuration composition of the states involved. Unfortunately, the interval between different energy levels in an ion also increases as $\sim (Z_i+1)^2$, which can quickly take the transition frequency out of the range of lasers as $Z_i$ increases. The phenomena of Coulomb degeneracy and configuration crossing can be used to combat this tendency. In a neutral atom, an electron orbital with a larger angular momentum is significantly higher than one with smaller angular momentum but with the same principal quantum number $n$. On the other hand, in the hydrogen-like limit orbitals with different angular momentum but the same principal quantum number are nearly degenerate. Therefore, somewhere in between there can be a crossing point where two levels with different angular momentum and principal quantum number can come close together: in such cases the excitation energy may be within laser range.

In Ref.~\cite{berengut10prl} we showed, using the Ag isoelectronic sequence as an example, why high $q$-values can occur in highly charged ions, and how the tendency of such systems towards large transition frequencies could be overcome. A two-valence-electron ion, Sm$^{14+}$, was identified, which has optical transitions that are the most sensitive to potential variation of $\alpha$ ever found. While atomic spectroscopy in electron beam ion traps is currently not competitive with optical frequency standards (see, e.g.,~\cite{draganic03prl,crespo08cjp} and review~\cite{beiersdorfer09pscr}) the technology continues to improve, and with the enhancements in sensitivity, highly-charged ions may prove to be a good system for detecting variation of $\alpha$.

\subsection{Enhanced effect of variation in UV transition of $^{229}$Th nucleus}
\label{sec:Th}

The $^{229}$Th nucleus has the lowest known excited state, lying just $7.6\pm 0.5$ eV above the ground state~\cite{beck07prl}. The position of this level was determined from the energy differences of many high-energy $\gamma$-transitions to the ground and first-excited states. The subtraction produces the large  uncertainty in the position of the 7.6 eV excited state. The width of this level is estimated to be about $10^{-4}$ Hz \cite{tkalya00prc}, which explains why it is so hard to find the direct radiation in this very weak transition. Nevertheless, the search for the direct radiation continues.

Because the $^{229}$Th transition is very narrow and can be investigated with laser spectroscopy, it is a possible reference for an optical clock of very high accuracy~\cite{peik03epl}. The near degeneracy of these isomers is a result of cancellation between very large energy contributions (order of MeV). Since these contributions would have different dependences on fundamental constants, this transition would be a very sensitive probe of possible variation of fundamental constants~\cite{flambaum06prl}. A rough estimate for the relative variation of the $^{229}$Th transition frequency is
\begin{equation}\label{deltaf}
	\frac{\delta \omega}{\omega} \approx 10^5 \left(
	0.1 \frac{\delta \alpha}{\alpha} +   \frac{\delta X_q}{X_q }\right)\,.
\end{equation}
Therefore, the experiment would have the potential of improving the sensitivity to temporal variation of the fundamental constants by many orders of magnitude.

More accurate nuclear calculations give different values for the sensitivity of this transition to $\alpha$. Refs.~\cite{hayes07plb,hayes08prc} claim that both isomers have identical deformations and therefore there is no enhancement of $\alpha$-variation. Other calculations give enhancement factors in the range $10^2$ -- $10^5$, depending on particulars of the model used~\cite{he07jpg,he08jpg,flambaum09epl,litvinova09prc}. To resolve this, we have proposed a method of extracting sensitivity to $\alpha$-variation using direct laboratory measurements of the change in nuclear mean-square charge radius between the isomers (isomeric shift)~\cite{berengut09prl}.
 
From \eref{deltaf}, we obtain the following energy shift in the 7.6 eV $^{229}$Th transition:
\begin{equation}\label{delta3}
\delta \omega \approx 
\frac{\delta X_q}{X_q}\ \textrm{MeV}\,.
\end{equation}
This corresponds to the frequency shift $\delta \nu \approx 3\cdot 10^{20}\,\delta X_q/X_q$ Hz. The width of this transition is $10^{-4}$ Hz so one may hope to get the sensitivity to the variation of $X_q$ about $10^{-24}$ per year. This is  $10^{10}$ times better than the current atomic clock limit on the variation of $X_q$.
 
Note that there are other narrow low-energy levels in nuclei, for example the 76~eV level in $^{235}$U with lifetime 26.6 minutes is the second-lowest known. One may expect a similar enhancement there. Unfortunately, this level cannot be reached with usual lasers. In principle, it may be investigated using a free-electron laser or synchrotron radiation. However, the accuracy of the frequency measurements is much lower in this case.

\ack
This work is supported by the Australian Research Council, Marsden grant and ECT*.

\bibliographystyle{iopart-num}

\begin{thebibliography}{10}
\expandafter\ifx\csname url\endcsname\relax
  \def\url#1{{\tt #1}}\fi
\expandafter\ifx\csname urlprefix\endcsname\relax\def\urlprefix{URL }\fi
\providecommand{\eprint}[2][]{\url{#2}}

\bibitem{webb10arxiv}
Webb J~K, King J~A, Murphy M~T, Flambaum V~V, Carswell R~F and Bainbridge M~B
  2010 ``Evidence for spatial variation of the fine structure constant''
  (\textit{Preprint} \eprint{arXiv:1008.3907})

\bibitem{flambaum04prd}
Flambaum V~V, Leinweber D~B, Thomas A~W and Young R~D 2004 {\em \prd\/} {\bf
  69} 115006

\bibitem{flambaum06prc}
Flambaum V~V and Tedesco A~F 2006 {\em \prc\/} {\bf 73} 055501

\bibitem{flambaum09ijmpa}
Flambaum V~V and Berengut J~C 2009 {\em \ijmpa\/} {\bf 24} 3342

\bibitem{marciano84prl}
Marciano W~J 1984 {\em \prl\/} {\bf 52} 489

\bibitem{langacker02plb}
Langacker P, Segr\`e G and Strassler M~J 2002 {\em \plb\/} {\bf 528} 121

\bibitem{calmet02epjc}
Calmet X and Fritzsch H 2002 {\em \epjc\/} {\bf 24} 639

\bibitem{wetterich03jcap}
Wetterich C 2003 {\em \jcap\/} {\bf JCAP10} 002

\bibitem{dent03npb}
Dent T and Fairbairn M 2003 {\em \npb\/} {\bf 653} 256

\bibitem{dmitriev04prd}
Dmitriev V~F, Flambaum V~V and Webb J~K 2004 {\em \prd\/} {\bf 69} 063506

\bibitem{berengut10plb}
Berengut J~C, Flambaum V~V and Dmitriev V~F 2010 {\em \plb\/} {\bf 683} 114

\bibitem{lea07rpp}
Lea S~N 2007 {\em \rpp\/} {\bf 70} 1473

\bibitem{savedoff56nat}
Savedoff M~P 1956 {\em \nat\/} {\bf 178} 689

\bibitem{uzan03rmp}
Uzan J~P 2003 {\em \rmp\/} {\bf 75} 403

\bibitem{dzuba99prl}
Dzuba V~A, Flambaum V~V and Webb J~K 1999 {\em \prl\/} {\bf 82} 888

\bibitem{dzuba99pra}
Dzuba V~A, Flambaum V~V and Webb J~K 1999 {\em \pra\/} {\bf 59} 230

\bibitem{dzuba03pra0}
Dzuba V~A, Flambaum V~V and Marchenko M~V 2003 {\em \pra\/} {\bf 68} 022506

\bibitem{dzuba05pra}
Dzuba V~A and Flambaum V~V 2005 {\em \pra\/} {\bf 71} 052509

\bibitem{dzuba02praA}
Dzuba V~A, Flambaum V~V, Kozlov M~G and Marchenko M 2002 {\em \pra\/} {\bf 66}
  022501

\bibitem{berengut04praB}
Berengut J~C, Dzuba V~A, Flambaum V~V and Marchenko M~V 2004 {\em \pra\/} {\bf
  70} 064101

\bibitem{berengut05pra}
Berengut J~C, Flambaum V~V and Kozlov M~G 2005 {\em \pra\/} {\bf 72} 044501

\bibitem{berengut06pra}
Berengut J~C, Flambaum V~V and Kozlov M~G 2006 {\em \pra\/} {\bf 73} 012504

\bibitem{dzuba08pra}
Dzuba V~A and Flambaum V~V 2008 {\em \pra\/} {\bf 77} 012514

\bibitem{berengut09mmsait}
Berengut J~C, Dzuba V~A, Flambaum V~V, King J~A, Kozlov M~G, Murphy M~T and
  Webb J~K 2009 {\em Mem. Soc. Astron. It.\/} {\bf 80} 795 (\textit{Preprint}
  \eprint{arXiv:physics/0408017})

\bibitem{webb99prl}
Webb J~K, Flambaum V~V, Churchill C~W, Drinkwater M~J and Barrow J~D 1999 {\em
  \prl\/} {\bf 82} 884

\bibitem{webb01prl}
Webb J~K, Murphy M~T, Flambaum V~V, Dzuba V~A, Barrow J~D, Churchill C~W,
  Prochaska J~X and Wolfe A~M 2001 {\em \prl\/} {\bf 87} 091301

\bibitem{murphy03mnras}
Murphy M~T, Webb J~K and Flambaum V~V 2003 {\em \mnras\/} {\bf 345} 609

\bibitem{srianand04prl}
Srianand R, Chand H, Petitjean P and Aracil B 2004 {\em \prl\/} {\bf 92}
  121302

\bibitem{murphy07prl}
Murphy M~T, Webb J~K and Flambaum V~V 2007 {\em \prl\/} {\bf 99} 239001

\bibitem{murphy08mnras}
Murphy M~T, Webb J~K and Flambaum V~V 2008 {\em \mnras\/} {\bf 384} 1053

\bibitem{srianand07prl}
Srianand R, Chand H, Petitjean P and Aracil B 2007 {\em \prl\/} {\bf 99} 239002

\bibitem{hinshaw09apjss}
Hinshaw G, Weiland J~L, Hill R~S, Odegard N, Larson D, Bennett C~L, Dunkley J,
  Gold B, Greason M~R, Jarosik N, Komatsu E, Nolta M~R, Page L, Spergel D~N,
  Wollack E, Halpern M, Kogut A, Limon M, Meyer S~S, Tucker G~S and Wright E~L
  2009 {\em \apjss\/} {\bf 180} 225

\bibitem{chand04aap}
Chand H, Srianand R, Petitjean P and Aracil B 2004 {\em \aap\/} {\bf 417}
  853

\bibitem{chand05aap}
Chand H, Petitjean P, Srianand R and Aracil B 2005 {\em \aap\/} {\bf 430}
  47

\bibitem{quast04aap}
Quast R, Reimers D and Levshakov S~A 2004 {\em \aap\/} {\bf 415} L7

\bibitem{levshakov05aap}
Levshakov S~A, Centuri\'on M, Molaro P and D'Odorico S 2005 {\em \aap\/} {\bf
  434} 827

\bibitem{levshakov06aap}
Levshakov S~A, Centuri\'on M, Molaro P, D'Odorico S, Reimers D, Quast R and
  Pollmann M 2006 {\em \aap\/} {\bf 449} 879

\bibitem{levshakov07aap}
Levshakov S~A, Molaro P, Lopez S, D'Odorico S, Centuri\'on M, Bonifacio P,
  Agafonova I~I and Reimers D 2007 {\em \aap\/} {\bf 466} 1077

\bibitem{reinhold06prl}
Reinhold E, Buning R, Hollenstein U, Ivanchik A, Petitjean P and Ubachs W 2006
  {\em \prl\/} {\bf 96} 151101

\bibitem{ubachs07jms}
Ubachs W, Buning R, Eikema K~S~E and Reinhold E 2007 {\em \jms\/} {\bf 241} 155

\bibitem{salumbides08prl}
Salumbides E~J, Bailly D, Khramov A, Wolf A~L, Eikema K~S~E, Vervloet M and
  Ubachs W 2008 {\em \prl\/} {\bf 101} 223001

\bibitem{king08prl}
King J~A, Webb J~K, Murphy M~T and Carswell R~F 2008 {\em \prl\/} {\bf 101}
  251304

\bibitem{malec10mnras}
Malec A~L, Buning R, Murphy M~T, Milutinovic N, Ellison S~L, Prochaska J~X,
  Kaper L, Tumlinson J, Carswell R~F and Ubachs W 2010 {\em \mnras\/} {\bf 403}
  1541

\bibitem{berengut10arxiv1}
Berengut J~C, Flambaum V~V, King J~A, Curran S~J and Webb J~K 2010 ``Is there
  further evidence for spatial variation of fundamental constants?''
  (\textit{Preprint} \eprint{arXiv:1009.0591})

\bibitem{tzanavaris05prl}
Tzanavaris P, Webb J~K, Murphy M~T, Flambaum V~V and Curran S~J 2005 {\em
  \prl\/} {\bf 95} 041301

\bibitem{tzanavaris07mnras}
Tzanavaris P, Webb J~K, Murphy M~T, Flambaum V~V and Curran S~J 2007 {\em
  \mnras\/} {\bf 374} 634

\bibitem{kanekar10apjlett}
Kanekar N, Prochaska J~X, Ellison S~L and Chengalur J~N 2010 {\em \apjlett\/}
  {\bf 712} L148

\bibitem{murphy01mnrasD}
Murphy M~T, Webb J~K, Flambaum V~V, Drinkwater M~J, Combes F and Wiklind T 2001
  {\em \mnras\/} {\bf 327} 1244

\bibitem{kanekar05prl}
Kanekar N, Carilli C~L, Langston G~I, Rocha G, Combes F, Subrahmanyan R, Stocke
  J~T, Menten K~M, Briggs F~H and Wiklind T 2005 {\em \prl\/} {\bf 95} 261301

\bibitem{kanekar10apjlett0}
Kanekar N, Chengalur J~N and Ghosh T 2010 {\em \apjlett\/} {\bf 716} L23

\bibitem{murphy08sci}
Murphy M~T, Flambaum V~V, Muller S and Henkel C 2008 {\em \sci\/} {\bf 320}
  1611

\bibitem{henkel09aap}
Henkel C, Menten K~M, Murphy M~T, Jethava N, Flambaum V~V, Braatz J~A, Muller
  S, Ott J and Mao R~Q 2009 {\em \aap\/} {\bf 500} 725

\bibitem{drinkwater98mnras}
Drinkwater M~J, Webb J~K, Barrow J~D and Flambaum V~V 1998 {\em \mnras\/} {\bf
  295} 457

\bibitem{van-veldhoven04epjd}
\mbox{van Veldhoven} J, K\"upper J, Bethlem H~L, Sartakov B, \mbox{van Roij}
  A~J~A and Meijer G 2004 {\em \epjd\/} {\bf 31} 337

\bibitem{henkel05aap}
Henkel C, Jethava N, Kraus A, Menten K~M, Carilli C~L, Grasshoff M, Lubowich D
  and Reid M~J 2005 {\em \aap\/} {\bf 440} 893

\bibitem{combes97apj}
Combes F and Wiklind T 1997 {\em \apj\/} {\bf 486} L79

\bibitem{flambaum07prl}
Flambaum V~V and Kozlov M~G 2007 {\em \prl\/} {\bf 98} 240801

\bibitem{dzuba00pra}
Dzuba V~A and Flambaum V~V 2000 {\em \pra\/} {\bf 61} 034502

\bibitem{angstmann04pra}
Angstmann E~J, Flambaum V~V and Karshenboim S~G 2004 {\em \pra\/} {\bf 70}
  044104

\bibitem{angstmann04pra0}
Angstmann E~J, Dzuba V~A and Flambaum V~V 2004 {\em \pra\/} {\bf 70} 014102

\bibitem{dzuba08pra0}
Dzuba V~A and Flambaum V~V 2008 {\em \pra\/} {\bf 77} 012515

\bibitem{flambaum09cjp}
Flambaum V~V and Dzuba V~A 2009 {\em \cjp\/} {\bf 87} 25 (\textit{Preprint}
  \eprint{arXiv:0805.0462})

\bibitem{berengut10arxiv0}
Berengut J~C and Flambaum V~V 2010 ``Manifestations of a spatial variation of
  fundamental constants on atomic clocks, Oklo, meteorites, and cosmological
  phenomena''
  (\textit{Preprint} \eprint{arXiv:1008.3957})

\bibitem{rosenband08sci}
Rosenband T, Hume D~B, Schmidt P~O, Chou C~W, Brusch A, Lorini L, Oskay W~H,
  Drullinger R~E, Fortier T~M, Stalnaker J~E, Diddams S~A, Swann W~C, Newbury
  N~R, Itano W~M, Wineland D~J and Bergquist J~C 2008 {\em \sci\/} {\bf 319}
  1808

\bibitem{nguyen04pra}
Nguyen A~T, Budker D, Lamoreaux S~K and Torgerson J~R 2004 {\em \pra\/} {\bf
  69} 022105

\bibitem{cingoz07prl}
Cing\"{o}z A, Lapierre A, Nguyen A~T, Leefer N, Budker D, Lamoreaux S~K and
  Torgerson J~R 2007 {\em \prl\/} {\bf 98} 040801

\bibitem{dzuba05pra0}
Dzuba V~A and Flambaum V~V 2005 {\em \pra\/} {\bf 72} 052514

\bibitem{angstmann06jpb}
Angstmann E~J, Dzuba V~A, Flambaum V~V, Karshenboim S~G and Nevsky
  A Yu 2006 {\em \jpb\/} {\bf 39} 1937

\bibitem{berengut10prl}
Berengut J~C, Dzuba V~A and Flambaum V~V 2010 {\em \prl\/} {\bf 105} 120801

\bibitem{draganic03prl}
Dragani\'{c} I, Crespo L\'opez-Urrutia J~R, DuBois R, Fritzsche S, Shabaev V~M,
  Orts R~S, Tupitsyn I~I, Zou Y and Ullrich J 2003 {\em \prl\/} {\bf 91} 183001

\bibitem{crespo08cjp}
\protect{Crespo L\'opez-Urrutia} J~R 2008 {\em \cjp\/} {\bf 86} 111

\bibitem{beiersdorfer09pscr}
Beiersdorfer P 2009 {\em \pscr\/} {\bf T134} 014010

\bibitem{beck07prl}
Beck B~R, Becker J~A, Beiersdorfer P, Brown G~V, Moody K~J, Wilhelmy J~B,
  Porter F~S, Kilbourne C~A and Kelley R~L 2007 {\em \prl\/} {\bf 98} 142501

\bibitem{tkalya00prc}
Tkalya E~V, Zherikhin A~N and Zhudov V~I 2000 {\em \prc\/} {\bf 61} 064308

\bibitem{peik03epl}
Peik E and Tamm \protect{Chr} 2003 {\em \epl\/} {\bf 61} 181

\bibitem{flambaum06prl}
Flambaum V~V 2006 {\em \prl\/} {\bf 97} 092502

\bibitem{hayes07plb}
Hayes A~C and Friar J~L 2007 {\em \plb\/} {\bf 650} 229

\bibitem{hayes08prc}
Hayes A~C, Friar J~L and M\"oller P 2008 {\em \prc\/} {\bf 78} 024311

\bibitem{he07jpg}
He \protect{X-t} and Ren \protect{Z-z} 2007 {\em \jpg\/} {\bf 34} 1611

\bibitem{he08jpg}
He \protect{X-t} and Ren \protect{Z-z} 2008 {\em \jpg\/} {\bf 35} 035106

\bibitem{flambaum09epl}
Flambaum V~V, Auerbach N and Dmitriev V~F 2009 {\em \epl\/} {\bf 85} 50005

\bibitem{litvinova09prc}
Litvinova E, Feldmeier H, Dobaczewski J and Flambaum V~V 2009 {\em \prc\/} {\bf
  79} 064303

\bibitem{berengut09prl}
Berengut J~C, Dzuba V~A, Flambaum V~V and Porsev S~G 2009 {\em \prl\/} {\bf
  102} 210801

\end{thebibliography}

\providecommand{\newblock}{}

\end{document}